\documentclass[runningheads]{llncs}
\usepackage{graphicx}
\usepackage{amsmath}
\usepackage{amssymb}
\usepackage{color,soul}
\usepackage{mathtools} 

\usepackage{enumitem}
\usepackage[colorinlistoftodos]{todonotes}

\newcommand{\eq}[1] {eq.\,(\ref{#1})}

\newcommand{\Eqs}[2]{Eqs.\,(\ref{#1}) and (\ref{#2})}

\newcommand{\singlequote}[1]{`#1'}

\newcommand{\dilate}{\delta}
\newcommand{\erode}{\varepsilon}
\newcommand{\open}{\gamma}
\newcommand{\close}{\phi}

\newcommand{\EqAnd}{\quad \text{and} \quad}

\newcommand{\bnd}{\overline{\partial}\,}
\newcommand{\com}{c\,}
\newcommand{\bcom}{\overline{c}\,}
\newcommand{\bdil}{\overline{\dilate}\,}
\newcommand{\bero}{\overline{\erode}\,}
\newcommand{\bopen}{\overline{\gamma}\,}
\newcommand{\bclose}{\overline{\phi}\,}

\newcommand{\bOpen}{\overline{\gamma}}
\newcommand{\bClose}{\overline{\phi}}
\newcommand{\bDil}{\overline{\dilate}}
\newcommand{\bEro}{\overline{\erode}}

 \newcommand{\half}{{\mbox{$\frac{1}{2}$}}}

\begin{document}
%

\title{Complementarity-Preserving Fracture Morphology for Archaeological Fragments\thanks{This research was funded by the GRAVITATE project under EU2020-REFLECTIVE-7-2014 Research and Innovation Action, grant no. 665155.}}
\titlerunning{Complementarity-Preserving Fracture Morphology}
%
\author{Hanan ElNaghy \and
Leo Dorst }
\authorrunning{H. ElNaghy and L. Dorst}
%
\institute{Computer Vision Lab, University of Amsterdam, The Netherlands
\email{\{hanan.elnaghy,l.dorst\}@uva.nl}}
\maketitle              
\begin{abstract}
We propose to employ scale spaces of mathematical morphology to hierarchically simplify fracture surfaces of complementarity fitting archaeological fragments.
This representation preserves complementarity and is insensitive to different kinds of abrasion affecting the exact fitting of the original fragments.
We present a pipeline for morphologically simplifying fracture surfaces, based on their Lipschitz nature;
its core is a new embedding of fracture surfaces to simultaneously compute both closing and opening morphological operations, using distance transforms. 
\end{abstract}
\section{Introduction}
The GRAVITATE H2020 project aims at providing archaeologists with the virtual tools to analyse digital artefacts, distributed across several collections (\url{https://gravitate-project.eu/}).
The artefacts are digitally scanned by standard scanning techniques in the form of 3D meshes, capturing the geometrical properties of each object and some of its photometric proprieties.
Reassembly of broken artefacts is one of the core objectives of the project.

The test case is on terracotta, which does not deform; but typical fragments are lacking material through abrasion and chipping.
Each fragment undergoes a preliminary preprocessing step \cite{ElNaghy_2017_ICCV} which partitions its surface into significant sub-parts called \singlequote{\emph{facets}}.
Each facet is characterized by its own geometrical properties of roughness of its surface and sharpness of its boundary, which in turn guides its categorization as either belonging to the fracture region or outside skin region of the fragment.

Our contribution to the project focuses on structuring the pair-wise alignment of promising fragments nominated for fitting by other selection modules.
The computational approach is based purely on the geometrical properties of the fracture facets, ignoring other clues (such as possible continuity of decorative patterns).
  \begin{figure*}
  \begin{center}
  \includegraphics[width=120 mm]{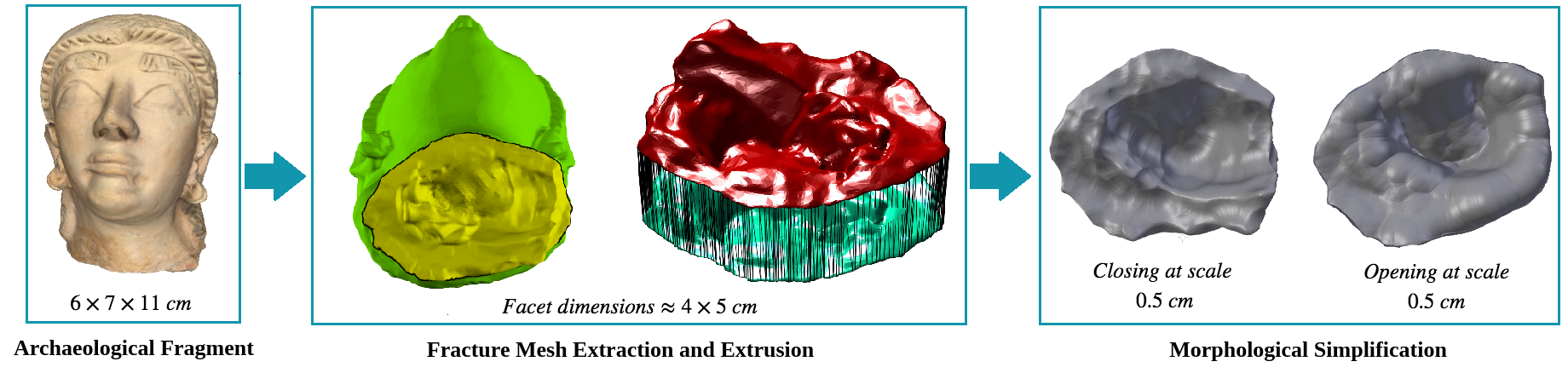}
        \caption {Morphological simplification pipeline for archaeological fracture facets.}
        \label{fig:fig_pipeline}
  \end{center}
  \end{figure*}
 
Optimal pair-wise alignment of potentially counter-fitting fragments is computationally expensive.
We therefore take a hierarchical approach, considering the fragments in increasing resolution from simplified to detailed.

The common way of simplifying shapes represented as meshes is by various forms of linear filtering of their vertices; letting them move with a differentiable flow (such as based on the Poisson equation) to produce a smoother model \cite{poisson}. 
Structure-aware mesh decimation to preserve the structure of the mesh by representing it as a set of pre-computed canonical proxies has been also studied by \cite{Decimation}.
Such simplifications are strongly related to characterizing a surface by differential geometric features, such as those computed by some means of discrete differential geometry \cite{crane}.
Indeed most methods to reassemble broken objects \cite{Pottmann,Fresco,Assembling,LearningFresco,CurveMatching} are all based on such features.
However, a small chip, resulting in a missing part of an object, will affect such differential measures in a {\em linear} fashion: the deeper the missing bit, the stronger the simplified shape is affected.
This is undesirable: the actual match between part and counterpart is simply lacking some local evidence of a more binary nature. The effect of this gap should not depend on the depth of the hole (a gap with the same surface outline but twice as deep is not twice as bad); so our representation method should not be too sensitive to this either.
For the same reason, applying ICP (Iterative Closest Point) algorithm to find a matching counterpart based on squared errors is not appropriate; some extensions of ICP that include distance transforms \cite{fitzgibbon} are more in line of what we might employ.
 
In our hierarchical approach, we need a representation where a coarser, simplified version of the fracture surface maintains its potential fit with a coarser, simplified version of the counterpart, under abrasion-type noise.
Mathematical morphology with its opening and closing scale spaces, of increasing size, is the natural choice for just that kind of representation.
Such scale spaces have been studied in \cite{jackway,boomgaardMM,bosworth}, and it has been demonstrated that they simplify the shape, when measured in terms of certain descriptive features such as local maxima \cite{jackway}, or zero-crossings of the curvature \cite{chenyan}.
But our intended use for alignment and complementarity checking appears to be new.
 
In this paper, we therefore hone standard Mathematical Morphology (MM) to this novel application.
First, we define scale-based complementarity of objects (rather than one object and its background). Then, we define morphological operations working only on the fracture region of the fragments.
We implement this \singlequote{Boundary Morphology} by taking advantage of the Lipschitz nature of the individual fracture facets.
Our extrusion method uses a single distance transform to simultaneously produce both scaled dilations and scaled erosions of a fracture facet (as in Figure~\ref{fig:fig_pipeline}), leading to a simplified representation at every scale.
We briefly discuss how the fracture Lipschitz property also naturally gives quantitative bounds on the detectable complementarity and collision-free alignment at each scale of our representation, even in the presence of abrasion.

 \section {Complementarity Preserving Scale Space}
 \label{sec_scale_space}
 Consider a fracture between two fragments $X$ and $Y$ in perfect alignment (or, if you prefer, consider a newly developed crack in an object).
 Within a well-chosen mask $M$ (such as the part of space the object occupies),
 the two fragments $X$ and $Y$ are almost complementary in the usual MM sense, with the fracture area being ambiguous (does it belong to $X$ or $Y$?).  In an implementation such an infinitely thin layer is not going to make a difference, so we arbitrarily choose $X$ to be a closed set and $Y$ to be open. Then the $M$-restricted parts of $X$ and $Y$, $X_M(=X\cap M)$ and $Y_M(=Y\cap M)$ are two relatively complementary sets $(X_M^c=Y_M)$.
 The properties of {\em exact complementarity} (no overlap, no gaps) can then be algebraically formulated as:
 \begin{eqnarray}
 \label{eq:1}
                 \text{No Overlap}&\Longleftrightarrow&\text{Non-Intersection: }X_M\cap Y_M=\emptyset \\
                  \label{eq:2}
                 \text{No Gaps}&\Longleftrightarrow&\text{Completeness: }X_M\cup Y_M = M
 \end{eqnarray}
 In archaeology, we do not acquire the fragments to be in alignment, or in perfect condition.
 We choose MM scale spaces to process the fracture region of each of the fragments separately, producing a simplified representation for each fracture facet at a range of scales $\rho$.
 Morphological scale spaces are well suited to this, since they can simplify objects while closely maintaining local complementarity, even when objects are slightly damaged (as we will discuss in Section \ref{sec_inexact}).
  
 Performing erosions or dilations would change the objects considerably, so we prefer to use \emph{openings} and \emph{closings} to process the fragments (even though much essential structure is already contained in eroded and dilated versions).  Because of the arbitrary orientation of presented objects, and an assumed isotropy of fracturing, our morphological scale space is built using structuring elements that are balls of radius $\rho$. 
  
 At a scale $\rho$, the opened version $\open_\rho(X)$ will be complementary (in the above sense) to the closed version $\close_\rho(Y)$, and vice versa.
 Since a closing $\close_\rho(X)$ simplifies the local geometry of the fracture at valleys, the complementarity with $\open_\rho(Y)$ becomes less specific in those areas (one can easily construct non-complementary counterparts $Y'$ with the same opening $\open_\rho(Y') = \open_\rho(Y)$, that do not fit the original $X$).
 To maintain specificity of peaks and valleys of the common fracture surface, we must therefore compute both $\open_\rho(X)$ and $\close_\rho(X)$  for each fragment $X$.
 Since both opening and closing are increasing, coarser levels of the scale space are guaranteed to contain less 
  detail, thus enabling the hierarchical approach (as well as being a justification for calling the process a \singlequote{simplification}).
  
  In a schema, if $R > \rho$, we have the following complementarity and containment relationships:
   \begin{equation}
    \label{eq:schema}
    \begin{aligned}
                 &\open_{R}(X_M) \subseteq \open_{\rho}(X_M)\subseteq X_M \subseteq \close_{\rho}(X_M) \subseteq \close_{R}(X_M)
                 \\& \quad\Updownarrow c ~~\quad ~~~~\Updownarrow c\quad  ~~~~\Updownarrow c
                 \quad ~~~~\Updownarrow c
                 \quad   ~~~~~~\Updownarrow c
                 \\&~\close_{R}(Y_M) \supseteq\close_{\rho}(Y_M) \supseteq Y_M ~\supseteq ~ \open_{\rho}(Y_M)\supseteq \open_{R}(Y_M)
    \end{aligned}
    \end{equation}
 Here $c$ refers to complementarity within a well-chosen mask $M$ containing the common fracture of $X$ and $Y$.
 This \singlequote{well-chosen} actually hides some essential details, since masking does not commute with morphology (e.g., $\close_\rho(X_M) \neq (\close_\rho(X))_M)$. It is not hard to show that when originally there was exact complementarity between $X$ and $Y$ within a mask $M$, then in the scale space exact complementarity  still holds between $\close_\rho(X)$ and $\open_\rho(Y)$, or between $\open_\rho(X)$ and $\close_\rho(Y)$, but is only guaranteed within a doubly eroded mask $\erode_{2\rho}(M)$.
    \section{Fracture Morphology}
     \label{sec_Fracture_Morphology}
   Consider a broken vase producing thin sherds as fragments.
    Volumetric morphological simplification of the fragments would be limited in scale by their thickness: an opening by a ball with a radius $\rho$ of more than half this thickness would results in empty sets, which would only be complementary to their counterpart in a trivial way.
   Since the fractures themselves have informative morphological structure transcending this scale, we design in this section a way to focus the morphological processing purely on the fracture surface (rather than on fragment volumes).
   This should suffice for our puzzle, since whether two broken fragments can be refit locally depends on the complementary shape of their fractures only.

%
%
 \subsection {Boundary Morphology}
  \label{sec_Boundary MM}
  In Section \ref{sec_scale_space}, we had masked objects $X_M$ and $Y_M$, with complementarity in $M$ (and a slight issue of how we treat their common boundary fracture $F$).
  Now consider only one of them, no longer in contact, and investigate how to compute its openings and closings.
  We are only interested in the effect on $F$, not on the remainder of the object volume within $M$.
  
  %

  $F$ is a boundary, not a volume, and the classical MM does not apply immediately.
  We first define what we mean by applying mathematical morphology to a boundary.
  Let us call it \emph{\singlequote{Boundary Morphology}} and denote its operations by an over-bar.
  Consider a general object $A$ of which we take the boundary $\partial A$ (and ignore the masking effects for the moment).
  As we have seen, for true complementarity some objects may be open sets, others closed sets.
  We need to take the boundary of either, so we will always close the set and then apply the usual boundary  operator $\partial$ to define our boundary operator (which we denote by $\bnd$):
  \begin{equation}
  \label{eq:8}
  \begin{aligned}
  	&\bnd A = \partial \overline{A}
  \end{aligned}
  \end{equation}
Thus $\bnd A$ returns the set of points of $\overline{A}$ that have neighbours in ${\overline{A}}^c$. 
  
  In order to distinguish erosion and dilation, we need to make the boundary oriented, for instance by denoting the outward pointing normal at each location.
  Then this orientation changes to its opposite if we consider the boundary of the complement.
  To more easily denote this, write the complement as the prefix operator $\com$.
  We then have:
  \begin{equation}
 \label{eq:9}
  \begin{aligned}
  	&\bcom \bnd A \equiv \bnd \com A ~(=\bnd A^c),
  \end{aligned}
  \end{equation}
  which defines the complementation $\bcom$ of the boundary.
  
  We define what it means to dilate and erode a boundary, defining operators $\bdil$ and $\bero$ in terms of classical volumetric operators by:
  \begin{equation}
 \label{eq:10}
  \begin{aligned}
  	& \bdil \bnd A \equiv \bnd \dilate A \EqAnd \bero \bnd A \equiv \bnd \erode A.
  \end{aligned}
  \end{equation}
  It now follows that we can perform an erosion as a dilation on the complemented (oppositely oriented) boundary:
  \begin{equation}
  \label{eq:11}
  \begin{aligned}
  	&  \bero \bnd A = \bnd \erode A = \bnd \com \dilate \com A = \bcom \bnd \dilate \com A = \bcom \bdil \bnd \com A = \bcom \bdil \bcom \bnd A.
  \end{aligned}
  \end{equation}
  and accordingly,
   \begin{equation}
     \label{eq:12}
     \begin{aligned}
     	& \bero = \bcom\bdil \bcom \EqAnd \bdil  = \bcom\bero \bcom.
     \end{aligned}
     \end{equation}
  In fact, this is merely the duality between dilation and erosion, extended to their boundary versions (with $\overline{c}$ playing the role of complementation).
  So once we can dilate a boundary, we can use that both to produce the dilation $\bdil$ and the erosion $\bero$ (by doing dilation on $\bcom \bnd A$, the oppositely oriented boundary):
 
  This can be easily extended to boundary closing $\bclose$ and opening $\bopen$ defined as:
  \begin{equation}
   \label{eq:13}
   \begin{aligned}
   	&\bclose \bnd A \equiv \bero\bdil\bnd A \EqAnd \bopen \bnd A \equiv\bdil \bero\bnd A,
   \end{aligned}
   \end{equation}
for it follows from \Eqs{eq:11}{eq:12} that we can also perform an opening as a closing on the complemented boundary:
    \begin{equation}
     \label{eq:14}
     \begin{aligned}
     	& \bopen\bnd A =\bdil \bero\bnd A = \bcom\bero\bcom\bero\bnd A =\bcom\bero\bdil\bcom\bnd A = \bcom\bclose\bcom\bnd A. 
     \end{aligned}
     \end{equation}
    and accordingly,
       \begin{equation}
         \label{eq:15}
         \begin{aligned}
         	& \bopen  = \bcom\bclose\bcom \EqAnd \bclose   =\bcom\bopen\bcom.
         \end{aligned}
         \end{equation}
  
  Now, the schema of eq.\,(\ref{eq:schema}) can be rewritten to describe the complementarity of two exactly fitting fracture boundaries (at scales $R$ and $\rho$ with $R>\rho$), as follows:
     \begin{equation}
     \label{eq:16}
     \begin{aligned}
     &\bOpen_{R}\bnd(X_M) \subseteq \bOpen_{\rho}\bnd(X_M)\subseteq \bnd(X_M) \subseteq \bClose_{\rho}\bnd(X_M) \subseteq \bClose_{R}\bnd(X_M)
                      \\&~ \quad\Updownarrow \bcom ~~~~~\quad ~~~~\Updownarrow \bcom\quad  ~~~~~\Updownarrow \bcom
                      \quad ~~~~~~~~\Updownarrow \bcom
                      \quad  ~ ~~~~~~~\Updownarrow \bcom
                      \\& ~\bClose_{R}\bnd(Y_M) \supseteq~\bClose_{\rho}\bnd(Y_M) \supseteq \bnd(Y_M) \supseteq~ \bOpen_{\rho}\bnd(Y_M)\supseteq \bOpen_{R}\bnd(Y_M)                      
     \end{aligned}
     \end{equation} 
     where $\bcom$ (complementarity) is taken within $M$. 
  \subsection{Lipschitz Condition}
     \label{sec_Lipschitz}
        \begin{figure*}
        \begin{center}
        \includegraphics[width=114.9 mm]{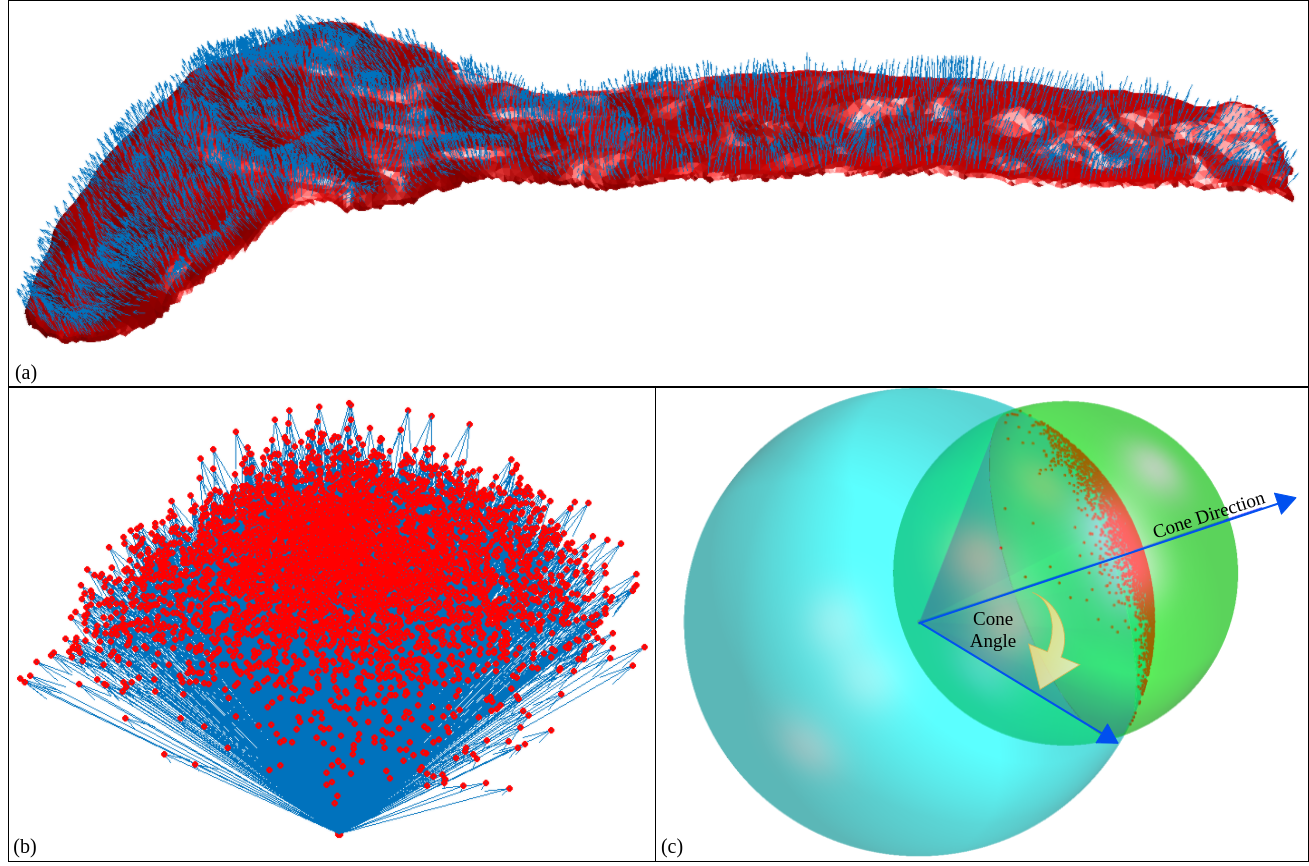}
              \caption {(a) Fracture surface with blue local normal directions. 
                            (b) Mapping of the normals to a unit sphere of directions.
                            (c) The minimum bounding sphere (in green) determines the minimum bounding cone on the unit sphere of directions (in cyan).}
              \label{fig:fig_Lipschitz}
        \end{center}
        \end{figure*}
  Terracotta is uniform and brittle.
  It fractures rather simply such that there is at least one 3D direction from which the whole fracture facet is completely visible, i.e. a ray in that direction emanating from each point of the fracture will not encounter any other point of that local fracture facet.
 Accordingly, a given fracture facet can be represented as a function (a Monge patch); noting that another fracture facet of the same fragment may require a different visibility direction to be seen as a function.
  The visibility assumption can be efficiently characterized in terms of the \singlequote{Lipschitz condition} which frequently occurs in quantitative mathematical morphology \cite{Lipschitz}. 
  
  For any pair of points on the graph of a slope-limited Lipschitz function with slope $s$, the absolute value of the slope of the line connecting them is always less than $s$ (i.e., the Lipschitz constant).
  As a consequence, there exists a double cone whose vertex can be moved along any such Lipschitz-continuous function, so that it always remains entirely visible from the principal direction of the cone while the rest of the function is completely outside the cone.
  For a local fracture to meet the visibility condition, it should meet the Lipschitz condition over its entire domain.
  
   To compute the Lipschitz slope $s$, we first collect the local normal vectors of the fracture facet on a unit sphere of directions, and then fit the largest possible cone inside this set (see Figure \ref{fig:fig_Lipschitz}).
   The axis of this cone we call the \emph{cone direction}, and the opening angle we call the \emph{cone angle}.
  Let us use the cone direction as the principal direction to align the fracture facet, so that it can be described as a Lipschitz function.
  This fracture function is then bounded in Lipschitz slope $s$, which is the cotangent of the cone angle.

  In practice, the Lipschitz condition appears to intrinsically hold for each of our terracotta fractures.
  Even if it does not hold, we can always artificially separate the fracture regions by the \emph{Faceting} preprocessing procedure \cite{ElNaghy_2017_ICCV} that delivers the piecewise Lipschitz facets of the fracture region.
%
  
  \subsection{Extrusion}
   \label{sec_Extrusion}
   \begin{figure*}
   \begin{center}
    \includegraphics[width=120 mm]{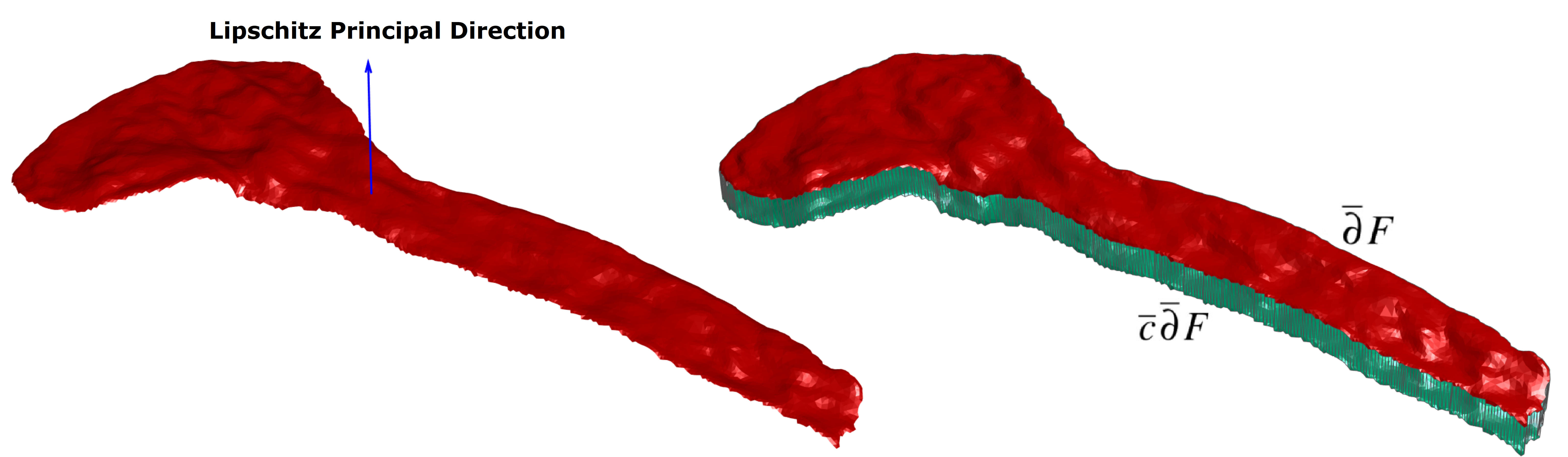}
         \caption {Extrusion of a fracture facet $\bnd F$ along the Lipschitz principal direction, generating a volume (cylinder)  with $\bnd F$ as its top red surface, and the oppositely oriented $\bcom \bnd F$ as its bottom aquamarine surface.}
         \label{fig:fig_extrsuion}
   \end{center}
   \end{figure*}
  The fracture surfaces, extracted using the Faceting preprocessing operation \cite{ElNaghy_2017_ICCV}, are locally Lipschitz.
  The corresponding Lipschitz principal direction acts as the \singlequote{average breakage direction} for the whole fracture facet.
  Therefore, we can generate a thickened fracture volume by extruding the facet along such visibility direction without self-intersection.
  Such an extrusion effect is equivalent to making two copies of the fracture surface bounded by a generalized cylinder: one with original (non-inverted) normals $\bnd F$ and one with inverted normals $\bcom \bnd F$(see Figure~\ref{fig:fig_extrsuion}).
  The outward propagation of the copy with non-inverted normals is the fracture surface dilation ($\bDil_\rho$), while the propagation of the one with inverted normals is the erosion ($\bEro_\rho$), by \eq{eq:12}.
  By a subsequent inward propagation of the resulting expanded surfaces, by the same amount $\rho$, one then acquires the closing ($\bClose_\rho$) and the opening ($\bOpen_\rho$) of the fracture.
  
   Focusing our morphological simplification on the extruded fracture volume has the following advantages:
   \begin{enumerate}[label=(\alph*)]
   \item Extending the functional morphological scale (ball radius $\rho$) beyond the minimum thickness of the archaeological object by only needing to do closing.
   \item Permitting simultaneous propagation of the fracture surface to both sides, thus allowing the production of MM opened and closed surfaces in one go.
   \item Avoiding the dilation of elements outside the mask (which could affect the outcome within the mask) by having the extruded surface surrounded by sufficient empty space completely isolated from outside influences.
   \item  Avoiding needless processing of the non-fracture regions of the object.
   \end{enumerate}
  

  
  \subsection{Distance-Transform based MM implementation}
    \label{sec_DistanceTransform}
     \begin{figure}
             \begin{center}
             \includegraphics[width=85 mm]{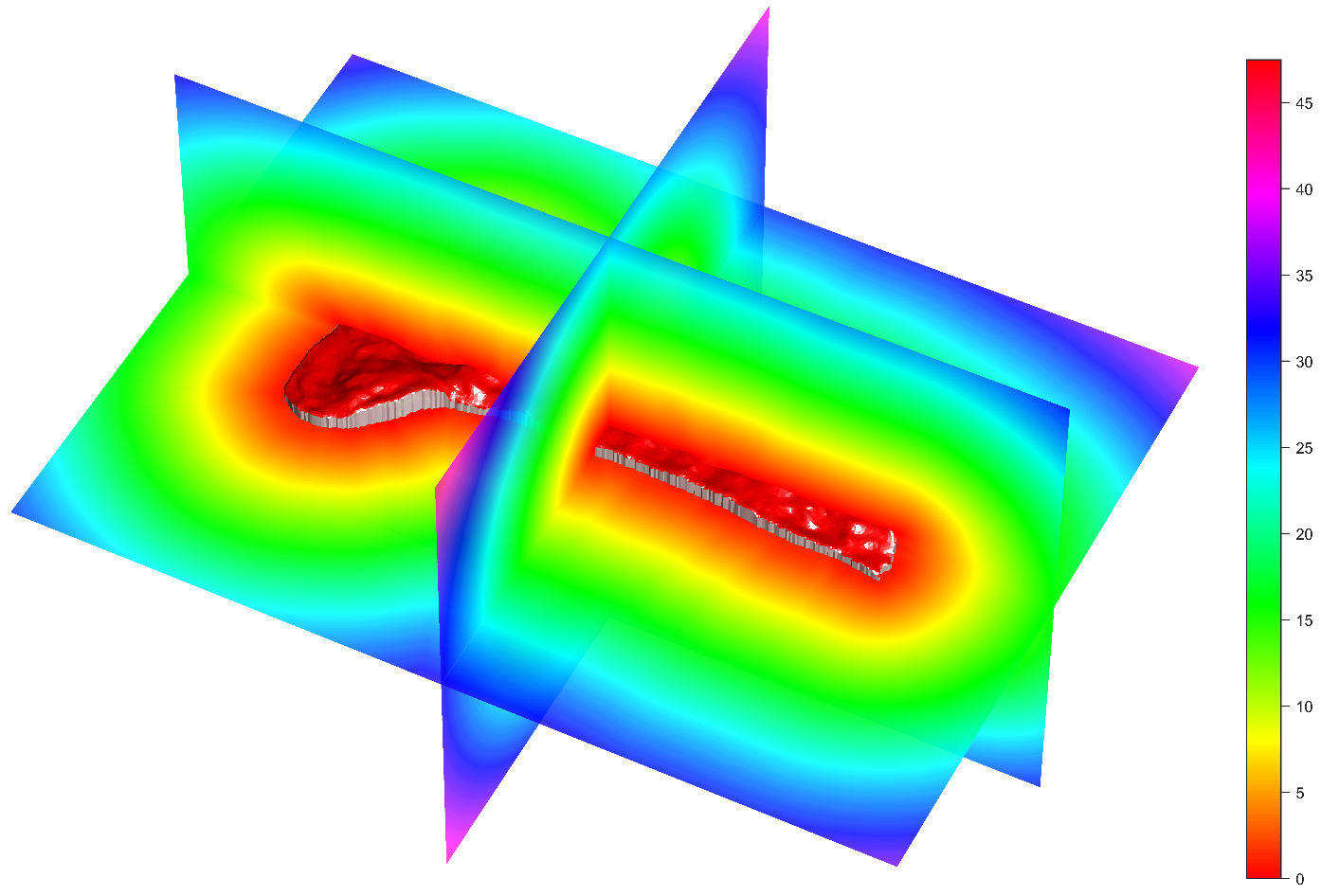}
                   \caption {Distance field computation, shown two cross sections}
                   \label{fig:fig_3DDF}
             \end{center}
             \end{figure}
   We obtain our data as 3D meshes, so it would be natural to apply the operations of boundary morphology to fracture facets represented directly as extruded 3D meshes.
   However, applying morphological operations directly on a mesh representation is notoriously hard (the dilation and erosion operations lead to many additional vertices as the mesh faces start intersecting \cite{meshMM}, and the ball structuring elements add many more).
   At this point of our narrative, it would seem that Point Morphology \cite{pointMorphology} is also a good candidate; we will see in Section \ref{sec_inexact} that we need more than just the simplified surfaces by themselves to determine our morphological features.
             
   An alternative implementation of morphology, especially when done in a scale space context, is by means of the isosurfaces of distance functions.
   Since we do want to perform our operations at different scales of a ball-shaped structuring element, this is especially attractive.
   Two basic approaches exist: distance transform considered as a numerical level set in a space of 1 more dimension \cite{sethian1998fast} or classic distance transform on a 3D grid.
   We adopted the latter approach which: a) suffices as a straightforward demonstration of the validity of our concepts and b) enables us to keep track of the original fracture points propagation in the distance field and their contribution to the generated MM surfaces through what we call the provenance map.

   Our implementation relies on distance transforms in a volumetric representation \cite{voxelization}, oriented with the Lipschitz principal direction along its z-direction.
   We convert the extruded mesh to binary representation by embedding the fracture surface in a 3D voxelized grid of a well chosen resolution, of step size $g$.
   We derived that the maximum outward and inward displacement that could take place due to discretization effects is no more than $\sqrt{3}g/2$.
   This implies that distances (and hence MM surfaces) are affected by no more than this amount.
   The generated grid is padded with an empty region sufficient to contain the maximally dilated versions of the fracture surfaces.
   For generating our distance field, the method proposed by \cite{Maurer03alinear} is employed, which calculates the euclidean distance transform in linear time on a binary voxelized representation of the object as shown in Figure \ref{fig:fig_3DDF}. 
   The closing of the fracture volume with a ball of radius $\rho$ (closing is all we need by virtue of \eq{eq:15}) is performed by first extracting the outward level set at distance $\rho$, then re-computing the distance field of the background and extracting the inward level set at distance ${\rho}$.
    \subsection{MM Surfaces}
    \label{sec_results}
     \begin{figure*}
                      \begin{center}
                      \includegraphics[width=120 mm]{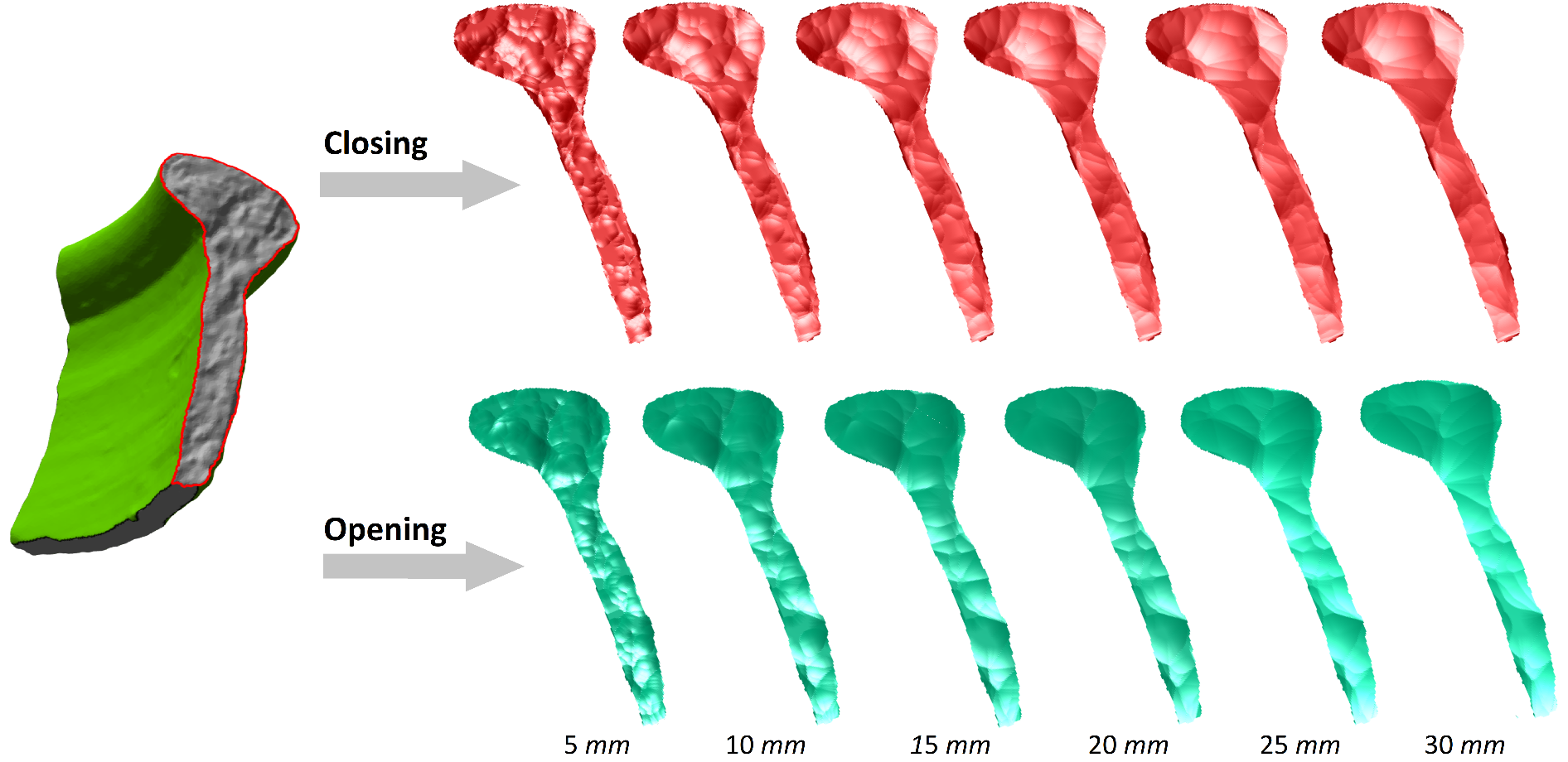}
                            \caption {Closing and opening scale spaces simplifying a fracture with increasing scale $\rho$.}
                            \label{fig:fig_closing_opening}
                      \end{center}
                      \end{figure*}
    Intuitively, morphologically closing the entire extruded volume is equivalent to rolling a ball on the fracture surface from both sides.
        The upper rolling (constrained by the fracture peaks) is equivalent to the closing effect, while the lower rolling (constrained by the fracture valleys) is equivalent to the desired opening effect.
               
    Figure \ref{fig:fig_closing_opening} shows the closed and opened simplified MM surfaces at 6 different scales of a given fracture surface with $5.8K$ vertices and $11K$ faces.
    The grid resolution is $g=0.2$ mm of size $119\times323\times29$ and is padded with $151$ voxels from each side to permit the outward propagation of the fracture surface up to $30$ mm without losing information, thus producing a grid of size $421\times625\times331$.
        
    
    The fracture Lipschitz computations together with the extrusion and embedding take less than $2$ seconds for a mesh of size $10K$.
    The distance field computation time is more affected by the resolution of the grid: computing the distance transform for $500\times500\times500$ grid takes at most 1 minute.
    Extracting the closed and opened MM surfaces takes less than $50$ seconds for the same grid size.
    In the GRAVITATE system, such times are considered acceptable.
    We performed our experiments on an Intel(R) Xeon(R) CPU $2.9GHZ$ $\times$ $8$ computer with $256GB$ RAM.
    
 \section {Inexact Complementarity}
 \label{sec_inexact}
  In practice, there are many aspects to our data that prevents complementarity from being exact.
  However, using the Lipschitz condition for fracture surface characterization together with MM for hierarchical fracture simplification enables us to establish the bounds on the amount of inexactness.
 In this section, we briefly discuss the sources causing inexactness and and how to treat their associated deviations.

 \begin{itemize}
  \item {\em \textbf{Abrasion}.}
  When a fractal surface as brittle as terracotta is rubbed, this will tend to
  take off the sharp local peaks without affecting the valleys.
  We propose that a reasonable model of such effect is that the object
  undergoes a morphological opening $\open_\alpha$ of a small size $\alpha$.
   
  Assuming this, the opening scale space will be unaffected for $\rho > \alpha$, since $\open_{\rho}(\open_\alpha(X)) = \open_\rho(X)$ for $\rho \geq \alpha$.
  However, the closing {\em is} affected by the abrasion.
  Therefore, exact complementarity of $\open_\rho(X')$ and $\close_\rho(Y')$ of \eq{eq:schema} schema no longer holds for the abraded version $X' = \open_\alpha(X)$ and $Y' = \open_\alpha(Y)$, even if it did for the original $X$ and $Y$.
  Nevertheless, the difference is bounded for the Lipschitz functions we are considering as the fracture facets.
  A simple sketch of a local sphere of radius $\alpha$ capped by a Lipschitz cone of slope $s$ shows that the maximal deviation of a possible original surface and its $\alpha$-opening is $(\sqrt{1+s^2}-1) \simeq \half \alpha s^2$.
  Since $X'$ does not differ by more than $\half\alpha s^2$ from $X$, also $\close_\rho(X')$ will not differ from $\close_\rho(X)$ by more than this amount. With that in mind, we can still test the possibility of complementarity, to within this bound, of the opened and closed abraded fractures.
  
  \item {\em \textbf{Fracture Adjacency Effects}.}
  Each facet is delineated by an outer border (contour).
  That border defines the complementarity zone of its processed MM surfaces.
  In practice, there are two different kinds of facet borders: \emph{fracture-skin} border and \emph{fracture-fracture} border.
  The former is the contour's segment that is adjacent to a facet which belongs to the exterior skin of the original object.  
  By contrast, a fracture-fracture border occurs when the original fragment is further broken into sub-fragments, causing the adjacency of more than one fracture facet.

   If the facet has only fracture-skin borders, then complementarity with its counterpart still holds over the entire surfaces of its simplified MM versions across all scales.
  For facets with fracture-fracture borders, one should expect complementarity to hold within regions away from the border.
   But there is a scale-dependent zone adjacent to the fracture-fracture borders where the MM surfaces are no longer guaranteed to be complementary to those of its counterpart.
   In that zone, the representation of either part may have been affected by the secondary breaking. 
   The bounds we can obtain on this zone under the general assumptions we made (such as Lipschitz) are weak.
   We have found that we can much more specifically delineate it through analysing the provenance map of the distance transform.
   Our current work is on efficient representation of the simplified scale space surfaces, taking this effect into account.

 \item {\em\textbf{ Misalignment}.}
                 Complementarity only exists in perfect alignment of the two counterparts; any offset in translation or orientation will destroy it.
                 But this effect is not as boolean as \Eqs{eq:1}{eq:2} makes it appear: the complementarity is quantifiable by means of the MM scale spaces.
                 The distance transform method of producing the openings and closings at different scales can be employed to guide the alignment, from coarser to finer scales, with the separation between one surface's opening and closing at each scale giving an indication of reasonable bounds on the fine alignment at that scale.
 \end{itemize}

 \section{Conclusion and Future Work}
  \label{sec_Conclusion}
 We have presented morphological simplification of the fracture facets of archaeological fragments based on their scanned 3D mesh representation, as a preparatory phase for optimal pair-wise alignment.
  This is a problem that lends itself very well to treatment by mathematical morphology, since that framework provides complementarity-preserving simplification of shapes in a manner that is insensitive to the kind of missing information and abrasion that we expect in the archaeological fragments.
  We showed how to perform morphology on boundaries in Section \ref{sec_Boundary MM}.
 The assumption that each fracture facet is a Lipschitz function allowed us to set up the extrusion method in Section \ref{sec_Extrusion}.
 The duplicate internal and external copies of the local fracture facet surface
 enable computing the opening and closing scale spaces simultaneously, in a volumetric representation in Section \ref{sec_results}.
  
 The resulting surfaces are simpler, but still contain the essentials of complementarity.
 We are currently investigating the compact characterization of the MM  scale space surfaces by means of characteristic scale space medial axis points, computed directly from the distance transform and its provenance map (specifying which points are influential at each location).  This should allow the use of standard registration algorithms on those  morphologically relevant feature points only.
 Moreover, we plan to use the provenance map of the distance transform to automatically control the reliability of those representations when parts of the original fracture facet are missing.

%

\bibliographystyle{splncs04}
\bibliography{mybibliography}

%
%
%
%
%
%
%
%
%
\end{document}